# Impacts of PV Capacity Allocation Methods on Distribution Planning Studies

Asmaa Alrushoud and Ning Lu
Department of Electrical and Computer Engineering
North Carolina State University
Raleigh NC, USA
aalrush@ncsu.edu, nlu2@ncsu.edu

*Abstract*—This paper presents a new method for assessing the amount of photovoltaics that can be accommodated on a distribution feeder before disrupting the normal operational conditions, commonly referred to hosting capacity. An optimal-capacity-based (OCB) PV allocation method is proposed to evaluate PV hosting capacity. We fist use load allocation method to allocate realistic load profiles to each load node down to each household. Instead of randomly assigning the installed capacity of PV to each household, the optimal size of PV for each house is first calculated based on the annual load profiles. Different PV allocation methods for hosting capacity calculation are first compared using the IEEE 123-bus system as a benchmark. An actual distribution feeder in North Carolina area is used to validate the results in realistic distribution systems. The simulation results show that the impact of PV capacity allocation methods on hosting capacity assessment is significant. We also investigate the zonal allocation method for weak zone identification to address the cluster phenomena in technology diffusion.

## I. INTRODUCTION

Behind-the-meter (BTM) installation of photovoltaic (PV) in power distribution systems has been increasing rapidly in the recent years. When the penetration increases, adverse impacts such as more frequent large voltage variations, longer duration of over- or under- voltage events, greater voltage unbalances, and thermal loading of distribution transformers and lines can occur. Therefore, determining the amount of PV that a distribution feeder can accommodate without violating its normal operational conditions, namely PV hosting capacity study [1] is increasingly critical in distribution system planning studies.

Hosting capacity is normally feeder specific and can assessed by running quasi-static, continuing power flow studies [2, 3]. Utility feeder models are used to determine the feeder topology. Load allocation methods are used to determine the load profiles on each load node. Voltage, load flow, operation statistics of the voltage regulation devices are calculated to estimate the severity of violation events.

Although several criteria, such as voltage, current, power quality and protection, are used to evaluate PV hosting capacity, the widely used criterion is voltage violations [4, 5]. Thus, over-voltage criterion is adopted in this paper for evaluating hosting capacity because it is one of the major concerns of the utilities and an important limiting factor to how much PV generation can be supported on a feeder.

In this paper, the PV hosting capacity is calculated using the stochastic analysis framework developed by Electric Power Research Institute (EPRI) [1]. The stochastic analysis captures the uncertainty in the size and location of future PV installations by populating PV deployment scenarios in a random fashion. In EPRI studies, for each PV deployment, the size of PV system at each customer location is randomly drawn from the residential or commercial PV distribution curves, depending upon the customer type [6], and the location is randomly selected from the pool of all load nodes provided by the distribution feeder.

Due to the recent installation of smart meters into the residential sector, yearly or longer household profiles has become available describing its consumption characteristics. So, in our study, we established a smart meter data base to represent the typical household load profiles in a region. Then, based on the feeder head load profiles, which are normally available to the utilities through SCADA measurements, a load allocation procedure [7] is carried out. This allows us to assign a number of households' profiles down to each load node. Then, an optimal PV size selection process is used to determine the PV size for each household based on cost benefit studies. This will allow each household to select a PV system that minimizes their monthly bill. Thus, the randomness in PV capacity allocation method is replaced by the optimal PV capacity allocation method. This method assumes that the selection of a BTM PV system is a rational decision made by the household owners to maximize their benefit.

The primary contribution of this paper is to correlate the residential load characteristics with the rooftop PV capacity selection. We demonstrated that the distribution grid can host more PV systems if households can select PV systems based on their consumption patterns as well as retail rate structures. Also, we demonstrated that when randomly allocating PV and PV capacity throughout the grid, PV hosting capacity estimation tends to under-estimate the PV hosting capacities. In addition, in cluster-based diffusion of a new technology, customers' acceptance rate can be higher in one neighborhood than another [8]. Thus, zonal allocation methods are more realistic than randomly allocate PVs to the whole feeder. Therefore, we develop a zonal based analysis method to identify the weaker zones.

The remainder of this paper is organized as follows. Section II describes the optimal PV sizing method and the methodology followed to determine PV hosting capacity. PV hosting capacity simulation results for different PV allocation methods are presented in Section III as well as the zonal-based analysis. Section IV presents Summary and future work.

## II. METHODOLOGY

This section introduces the simulation setup and the modeling methods

### A. Feeder Setup and Load Data Preparation

In this study, IEEE 123-bus feeder (see Fig. 1) and an actual feeder (see Fig. 2) have been used. The IEEE 123-bus feeder has 91 residential load nodes at 4.16 kV voltage level and the real life distribution feeder has 48 residential load nodes at 24 kV voltage level. The original feeder data contains only a peak load at each node. Thus, a load allocation method introduced in [7] is applied to generate load profiles at each load node and determine the total number of houses at each node. In order to conduct a time-series study, every load node on the feeder needs to be modeled down to the residential home level at 1-minute resolution.

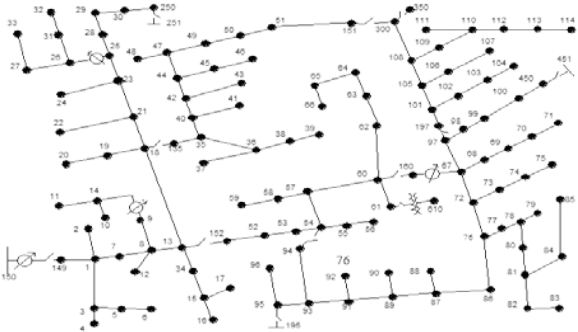

Fig. 1. IEEE 123-bus feeder topology

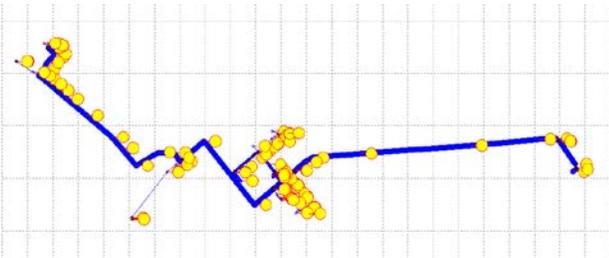

Fig. 2. Actual distribution feeder topology

A load profile database is established to construct nodal load profiles using two data sources. The first source is from Pecan Street website [9]. The data is minute by minute collected at Austin, Texas from July-August 2015. The second data source is provided by a utility at 30 minutes resolution. In Fig.3, a few examples of the house load profiles in our load pool are shown.

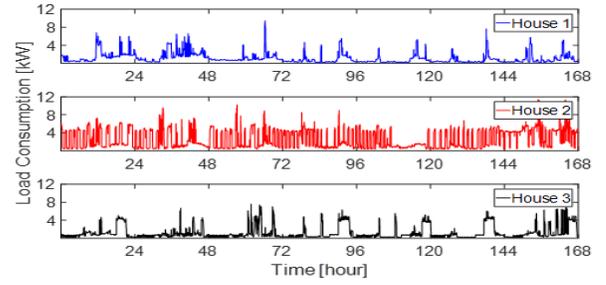

Fig. 3. Sample house load profiles

### B. Load Allocation Method

A bottom-up approach has been used to allocate weekly residential load profiles to every load node on a test feeder [7]. According to the data sheet, let the number of load nodes is $N_L$ and the peak load at the *ith* load node be $P_i$. For every load node, randomly draw weekly load profiles from the load pool and add them up until the aggregated load is bounded by $P_i$ upper and lower ranges as shown in Fig.4. The following stop criteria is used for the allocation process:

$$1.03 \times P_i \geq \max(\sum_{k=1}^{N_{i,k}} P_{i,k}) > 0.97 \times P_i \quad (1)$$
$$k \in \{1, \ldots, N_{i,k}\} \quad i \in \{1, \ldots, N_L\}$$

where $N_{i,k}$ is the number of houses at node $i$ and $P_{i,k}$ is the weekly load profile of house $k$ at load node $i$. By repeating the process for all load nodes in the test feeder, the test feeder has been prepared to perform time series studies and each load node has a number of houses allocated to it.

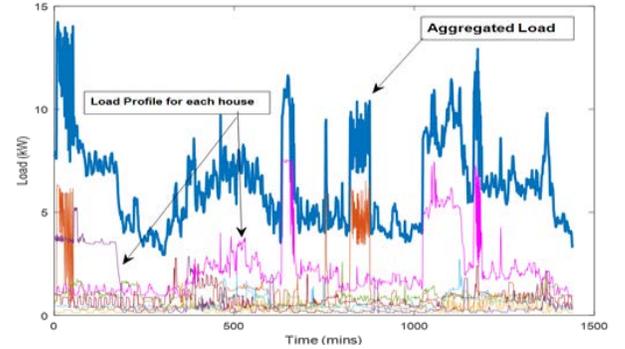

Fig. 4. Aggregated nodal load profile

### C. PV Optimal Sizing Selection Method

Because each household has its unique consumption characteristics, the house owner who wants to install PV on his rooftop will most likely consider optimizing the PV systems to either minimizing monthly bill or maximizing revenues when providing grid services. The wide deployment of smart meters makes household yearly load profiles available and accessible to homeowners. Thus, a cost-benefit based method has been conducted to find the optimal PV size for a specific house using yearly household load profiles and Time of Use (TOU) tariff of Duke Energy in North Carolina for the year 2018 [10] as inputs.

The TOU prices are shown in Table1. A demand charge of 13$/kW is considered. Optimality is defined as the PV size with the greatest annual net benefits, as shown in Fig.5. In this paper, the benefits is defined as the amount of savings in utility bills

between two cases, with and without PV under Duke's TOU rates and demand charges. We also assume that the back feeding PV power is not paid under this rate structure.

The levelized annual PV cost, $C$, can be expressed as

$$C = k_{PV}(a_{PV} \cdot P_{PV}) \quad (2)$$

$$k_{pv} = \frac{i(1+i)^y}{(1+i)^y - 1} \quad (3)$$

where $a_{PV}$ is the PV capital cost of $1000/kW and $k_{PV}$ capital recovery factor with lifetime $y = 20$ years and discount rate $i = 8\%$.

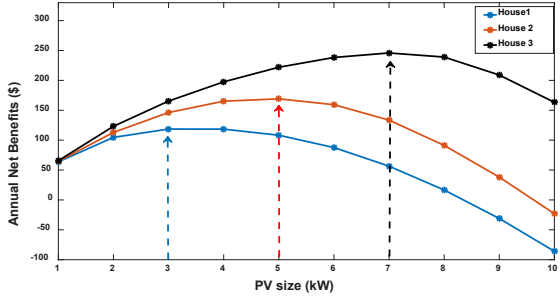

Fig. 5. Optimal PV sizes for three houses

TABLE I: Duke Energy ToU Rates

| Season | Load Type | Period | TOU($/kWh) |
|---|---|---|---|
| Summer (June-September) | Off-Peak | 9 PM -11AM | 0.07063 |
| | Partial-Peak | 11 AM -1 PM 6 PM -8 PM | 0.11996 |
| | Peak | 1 PM -6 PM | 0.23507 |
| Winter (October-May) | Off-Peak | 8 PM -6 PM 1 PM -5 PM | 0.07063 |
| | Partial-Peak | 9 AM -12 PM 5 PM -8 PM | 0.11708 |
| | Peak | 6 AM -9 AM | 0.22356 |

*D. PV Hosting Capacity*

PV Hosting capacity is defined as the maximum amount of PV that a feeder can accommodate before adverse impacts occur on a distribution feeder [1]. This value will be dependent on the feeder characteristics, PV size and location, monitoring criteria, and load. The overvoltage criterion is generally the primary concern of the power system utility and this is the criteria that will be used throughout the paper to determine PV hosting capacity. The overvoltage caused by PV output can be a major limiting factor to how much PV generation capacity can be supported on a distribution system. In order to find the PV hosting capacity, we utilized a modified version of the stochastic analysis framework that was developed by EPRI for determining the impacts of PV systems on a distribution circuit [1]-[3] as shown In Fig.6. The modification that was added is a step to determine the optimal PV size for each residential house prior performing the hosting capacity. Instead of randomly drawing PV sizes from distribution curves (residential/commercial), a unique and optimal PV size can be found for every house on every load node on a distribution feeder. Note that by selecting an optimal PV size for each household, the reverse flow of PV generation can be reduced.

The methodology to systematically simulate stochastic PV deployment scenarios is described in Fig. 6. We selected 100 PV deployment scenarios (i.e. $M = 100$). Each scenario is unique in the order that PVs are deployed and the maximum nodal voltage is recorded for each scenario by performing time series power flow. The location of the PV systems are also different in each scenario, which is the main random variable.

For a particular PV deployment scenario, the PV penetration level is increased from 0% - 100% in a step of 5% (i.e. $N = 20$). The PV penetration is the ratio between total PV installed capacity in a feeder (kWp) and the maximum apparent power on the feeder.

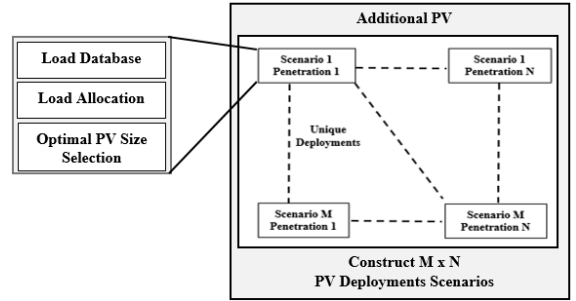

Fig. 6. Stochastic analysis framework

The procedure used to determine the PV hosting capacity is summarized as follows:

Step 1: Process load data sources to create load pool whether it is 1-mintue / 30-mintue resolution data.
Step 2: Perform load allocation method for all load nodes to match the feeder head load profile with the household load profiles in the load pool. This step will prepare the nodal load profiles for running quasi-static power flow simulations.
Step 3: Run optimal PV size selection algorithm for each household to determine the optimal PV size for each house on the feeder based on the net annual benefits calculated using the Duke Energy ToU prices.
Step 4: Perform stochastic analysis framework to determine PV hosting capacity using the voltage violation criterion.

III. SIMULATION RESULTS AND ANALYSIS

The analysis is performed in OpenDSS, an open-source system simulation software developed by EPRI [11]. OpenDSS is controlled through the COM interface by MATLAB. MATLAB is used for creating and iterating through each PV scenario as well as for the analysis of the results. OpenDSS is used to solve the time series power flow for each case. IEEE 123-bus feeder and an actual distribution feeder provided by a local utility were used in the simulation process.

*A. PV Hosting Capacity under multiple PV allocation*

The load allocation method applied to the IEEE 123-bus feeder have allocated 647 houses with weekly profiles to all load nodes aggregated to a substation load peak at 2.42 MW, and for the actual feeder it has 1573 houses with a substation load peak at 6.1 MW. Then, for every house, an optimal PV size was calculated based on its annual load profile. Stochastic analysis was performed to populate different PV deployments with optimal PV sizes and random locations starting from 5% PV penetration to 100%. In each PV deployment, a maximum bus voltage was recorded. The PV hosting capacity was calculated based on overvoltage criterion under multiple PV sizing methods. As shown in Figs.7-9, PV hosting capacity under optimal-capacity-based (OCB), randomly assigned, and standardized PV allocation methods are presented for the actual distribution feeder. In Fig.7 (the OCB case), the maximum voltage profile reaches a point where voltage violations decreases as the PV penetration increases until it settled to an acceptable voltage (with no violation) under 100% PV penetration with 6.1 MW of PV installed. As opposed to randomly assigned and standardized PV allocation methods, the maximum voltage rises as the PV penetration increases where it settled to unacceptable voltage under 100% PV penetration. Also, the minimum hosting capacity is 1.6 MW for randomly assigned and standardized PV allocation methods and it is 2.04 MW for OCB approach resulting in an increased capacity for PV without affecting the operational conditions.

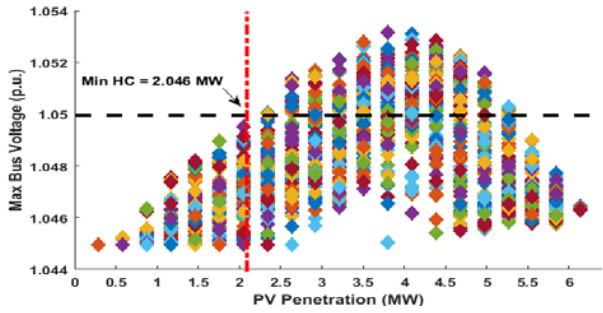

Fig. 7. Hosting capacity with OSB PV allocation

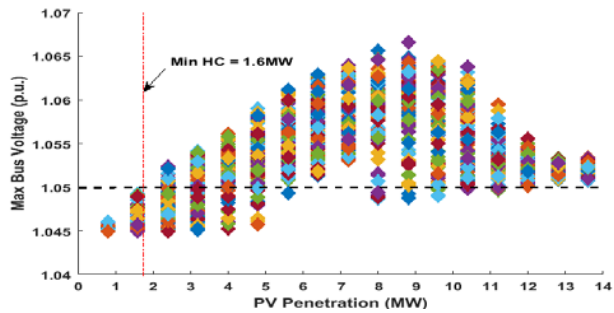

Fig. 8. Hosting capacity with random PV allocation

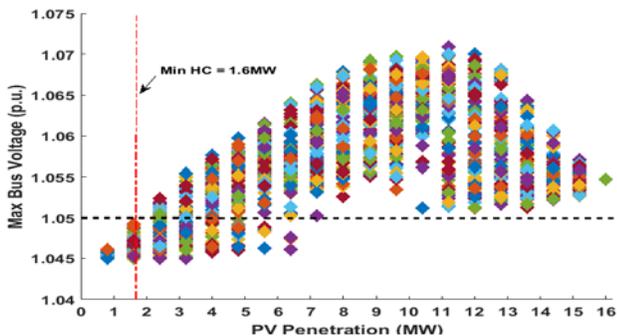

Fig. 9. Hosting capacity with standard 10 kW PV allocation

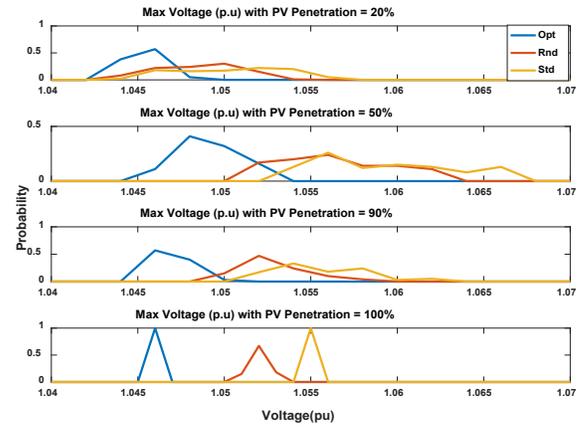

Fig. 10. Maximum voltage probability for optimal, random, 10 kW standard PV allocation methods

In Fig.10, a comparison between the three PV allocation methods under multiple PV penetration levels is presented. Here, the PV penetration is defined as the percentage of customers with PV systems. Thus, 100% PV penetration represents that all customers on the actual feeder has a PV system. It can be seen that when the PV penetration level is low, all PV allocation method have relatively close maximum voltage profiles. When the PV penetration level is high, the randomly assign and standardized PV capacity allocations will over-estimate the voltage violation limiting possible future PV installation as compared with the OCB approach.

### B. Zonal-Based Analysis

In a community, the installation of rooftop PV systems exhibits zonal characteristics because of the social contacts among neighbors. To model this phenomena in PV technology diffusion process, we applied a zonal PV allocation method in the distribution planning study. The IEEE 123-bus feeder is first divided into 10 zones using Proximity Analysis, as shown in Fig. 11. Note that the number of zones can vary depending on the feeder topology, load types, and the length of the feeder.

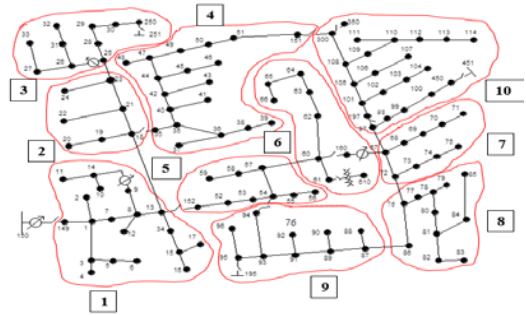

Fig. 11. Zones topology for IEEE 123-bus feeder

For each zone, 100% customers with PV systems is assumed and the voltage profile within that zone is recorded under different PV capacity allocations and then the voltage change between the case with no PV in the zone with the PV case is presented in Fig. 12.

From the results, we made the following observations:
- The OCB PV allocation method yields smaller voltage changes compared with random or 1-size PV allocation methods.

- Zonal voltage changes are correlated to electrical connection between zones and load characteristics. Criterions can be set up to identify the zones that can host more PVs or less PVs in order to increase the overall hosting capacity on the feeder.
- As expected, if a zone is located further away from the substation or a voltage regulator, voltage changes increases. Thus, we can use this method to identify the zones that are more vulnerable to voltage violations with concentrated PV installations.

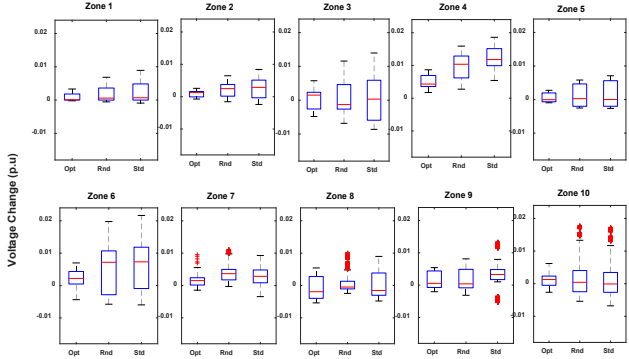

Fig. 12. Zonal voltage change (1-min data resolution)

*C. Load Data Sampling Rate Impacts*

As mentioned, two load data sources were used to perform load allocation method to generate load profiles at each load node. The data sampling rate from Pecan Street is one minute while the utility provided data is 30 minutes. The hosting capacity results under both data resolution in presented in Table2. With faster data sampling more benefits can be seen and full potential impacts can be captured. For zonal based analysis, the voltage changes for every zone is presented in Fig.13. It can be seen that for zones (7, 8, 9, and 10) the voltage changes is bigger as compared to Fig.12. It is because the control devices like voltage regulators bring the voltage level down in a short time span, so 1-min data or even faster is required to capture it accurately.

Table 2: Hosting Capacity for different PV allocation methods with different data sampling rate

| PV allocation methods | Min Hosting Capacity (MW) 1-min data | Min Hosting Capacity (MW) 30-min data |
|---|---|---|
| Optimal | 2.04 | 1.2 |
| Random | 1.6 | 0.9 |
| Standard (10 kW) | 1.6 | 0.9 |

IV. CONCLUSION

In this paper, we did not discuss the method for determining the optimal PV size and for separating the feeders into different zones due to the page limits. A follow-up journal paper will be written to introduce the detailed methodologies used in the study. Our simulation results demonstrate that it is of great importance for distribution planning engineers to apply optimal PV sizing and zonal allocation methods to identify the PV hosting capacity. Installing large PV systems in weaker zones will decrease the overall PV hosting capacity. Adding voltage regulation devices or letting customers choose an optimal size can alleviate the voltage violations.

When the PV penetration level is high, the ongoing PV capacity allocation methods, i.e. randomly assigning PV capacities or use a fixed PV capacity regardless or residential load characteristics will over-estimate the voltage violation and under-estimate the PV hosting capacity. Such approaches will limit future PV installation as compared with the OCB approach. High resolution end use data will facilitate the hosting capacity study by allowing more accurate estimations on voltage violation events and identifying optimal PV sizes for residential loads. Dividing a feeder into zones can also facilitate the PV hosting capacity study by addressing the clustering phenomena in the technology diffusion process. We expect the new considerations introduced in this paper will assist utility engineers improving PV hosting capacity studies and developing zonal based voltage control methods.

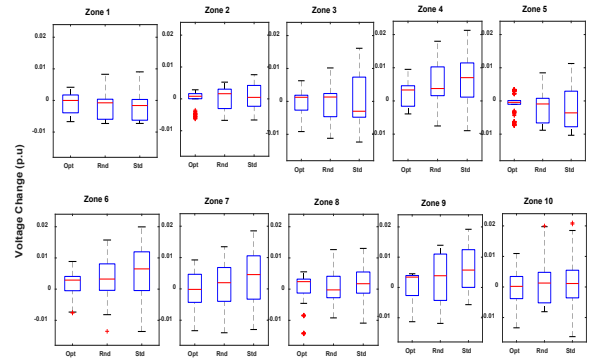

Fig. 13. Zonal voltage change (30-min data resolution)